\RequirePackage{fix-cm}
\documentclass[smallextended]{svjour3}       % onecolumn (second format)
\smartqed  % flush right qed marks, e.g. at end of proof
\usepackage{graphicx}
\begin{document}

\title{Electroexcitation of nucleon resonances in a
light-front relativistic quark model
}
\author{I. G. Aznauryan and V. D. Burkert 
}

\institute{I. G. Aznauryan \at
              Yerevan Physics Institute, 375036 Yerevan,Armenia \\
              Thomas Jefferson National Accelerator Facility,
Newport News, Virginia 23606, USA\\
              \email{aznaury@jlab.org}           %  \\
           \and
           V. D. Burkert \at
              Thomas Jefferson National Accelerator Facility,
Newport News, Virginia 23606, USA\\
              \email{burkert@jlab.org}           %  \\
}

\date{Received: date / Accepted: date}

\maketitle

\begin{abstract}
We report the predictions for the $3q$ core contributions
to the electroexcitation of the resonances
$\Delta(1232)\frac{3}{2}^+$,
$N(1440)\frac{1}{2}^+$,
$N(1520)\frac{3}{2}^-$, 
$N(1535)\frac{1}{2}^-$, and
$N(1675)\frac{5}{2}^-$
on the proton obtained in the
light-front relativistic quark model (LF RQM).
For these states, experimental data on the electroexcitation transition
amplitudes allow us to make comparison between the experiment
and LF RQM predictions in wide range of $Q^2$ and also to quantify
the expected meson-baryon contributions as a function of
$Q^2$.
\end{abstract}

\section{Introduction}
\label{intro}
Recently extensive investigation of the electroexcitation of nucleon resonances
has been made in the light-front dynamics, which is known as
most suitable framework for describing the transitions between relativistic
bound systems. In wide range of  $Q^2$, electroexcitation on the proton
and neutron is investigated in LF RQM \cite{Aznauryan1982}
for the resonances of the multiplets
$[56,0^+]$, $[56,0^+]_r$, and $[70,1^-]$
\cite{Aznauryan2007,Aznauryan2012,Aznauryan2015,Aznauryan2016,Aznauryan2017}.
In Refs. \cite{Aznauryan2012,Aznauryan2015,Aznauryan2016,Aznauryan2017},
the parameters of the LF RQM
have been specified via description of the nucleon electromagnetic form factors
up to $Q^2 = 16~$GeV$^2$ by combining the $3q$ and pion-cloud contributions
in the LF dynamics and by employing running quark mass as a function of $Q^2$
\cite{Aznauryan2012}.

In this talk we present the results for the transitions
$\gamma^* p \rightarrow \Delta(1232)\frac{3}{2}^+$,
$N(1440)\frac{1}{2}^+$,
$N(1520)\frac{3}{2}^-$, 
$N(1535)\frac{1}{2}^-$, and
$N(1675)\frac{5}{2}^-$. 
For these processes experimental data on the transition amplitudes
are available in wide range of $Q^2$ and, therefore, 
allow us to make detail comparison between the experiment
and LF RQM predictions and also to make conclusions on
the inferred meson-baryon contributions.

\section{Results and Discussion}
\label{results}

The predictions for the $\gamma^* N \rightarrow N$ form factors
and for the amplitudes of resonance electroexcitation on nucleons
have been made in Refs.
\cite{Aznauryan2012,Aznauryan2015,Aznauryan2016,Aznauryan2017}
under the assumption
that in addition to the three-quark (3q) contribution, these transitions
contain contributions, which are produced by meson-baryon
interaction.
The nucleon electromagnetic form factors were described
by combining $3q$ and pion-nucleon loops contributions.
With the pion-nucleon loops evaluated according to
the LF approach of Ref. \cite{Miller}, the following form of the nucleon
wave function has been found:
\begin{equation}
|N>=0.95|3q>+0.313|\pi N>.
\label{eq:eq1}
\end{equation}

For the resonances
$\Delta(1232)\frac{3}{2}^+$,
$N(1440)\frac{1}{2}^+$, $N(1520)\frac{3}{2}^-$, and
$N(1535)\frac{1}{2}^-$,
the weights of the $|3q>$ component in the expansion
\begin{equation}
|N^*>=c_{N^*}|3q>+...~~~
\label{eq:eq2}
\end{equation}
were found
from experimental data on
the $\gamma^* p \rightarrow \Delta(1232)\frac{3}{2}^+$,
$N(1440)\frac{1}{2}^+$, $N(1520)\frac{3}{2}^-$, and
$N(1535)\frac{1}{2}^-$ 
assuming that at $Q^2 > 2-3~$GeV$^2$
these transitions
are determined only by the first term in Eq. (\ref{eq:eq2}).
The obtained values of the coefficients $c_{N^*}$ are
$0.88\pm 0.04,~0.93\pm 0.05,~0.93\pm 0.06$, and $0.91\pm 0.03$,
respectively,
for the resonances
$\Delta(1232)\frac{3}{2}^+$,
$N(1440)\frac{1}{2}^+$, $N(1520)\frac{3}{2}^-$, and
$N(1535)\frac{1}{2}^-$. For the  $N(1675)\frac{5}{2}^-$,
we have good description of the data at  $Q^2 > 3.5~$GeV$^2$
with $c_{N^*}\simeq 1$; the presented results for the $3q$ core 
contribution to
the $\gamma^* p \rightarrow N(1675)\frac{5}{2}^-$ amplitudes
correspond to $c_{N^*}= 1$.

To study sensitivity to the form of the quark wave function,
two forms of the spatial wave function
have been employed in Ref.~\cite{Aznauryan2012}:
\begin{equation}
\Phi_1 \sim \exp(-M_0^2/6\alpha_1^2),
~~~~\Phi_2 \sim
\exp\left[-({\bf{q}}_a^2+{\bf{q}}_b^2+{\bf{q}}_c^2)/2\alpha_2^2\right],
\label{eq:res1}
\end{equation}
where $M_0$ is invariant mass of the system of constituent quarks,
and ${\bf{q}}_i$ are three momenta of these quarks in their c.m.s.
A good description
of the nucleon electromagnetic form factors up to $Q^2=16$~GeV$^2$
has been obtained 
with the wave functions (\ref{eq:res1})
using the following parameterizations for
the running quark mass as a function of $Q^2$:
\begin{equation}
m_q^{(1)}(Q^2)=\frac{0.22~{\rm GeV}}{1+Q^2/56~{\rm GeV}^2},
~~~m_q^{(2)}(Q^2)=\frac{0.22~{\rm GeV}}{1+Q^2/18~{\rm GeV}^2}.
\label{eq:res2}
\end{equation}

For the resonances, the results for
the transition amplitudes
obtained with the wave functions
(\ref{eq:res1})
and corresponding masses
(\ref{eq:res2})
are very close
to each other. The role of the running quark mass becomes visible
above $3$~GeV$^2$.
At $Q^2=5(12)$~GeV$^2$,
it increases the transition helicity amplitudes
by $25\%-35\%(100\%-140\%)$ and $10\%-15\%(55\%-65\%)$
for the  wave functions $\Phi_1$ and
$\Phi_2$, respectively.

We note that while in QCD lattice calculations and Dyson-Schwinger
equations we deal with the off-shell quarks, and the quark virtuality
is determined by their four-momentum square, in constituent quark models, 
including LF approaches, the quarks are on-mass-shell objects,
and their viruality is characterized by the invariant mass of the
quarks, which is incleasing with increasing $Q^2$.
The correspondence between $Q^2$ and mean value of $M_0^2$ is following:
$Q^2=0,~5,~10,~20~GeV^2$ correspond, respectively, 
to $<M_0^2>=1.35,~2.66,~3.1,~3.5~GeV^2$.

\begin{figure}
\includegraphics[width=5.7cm]{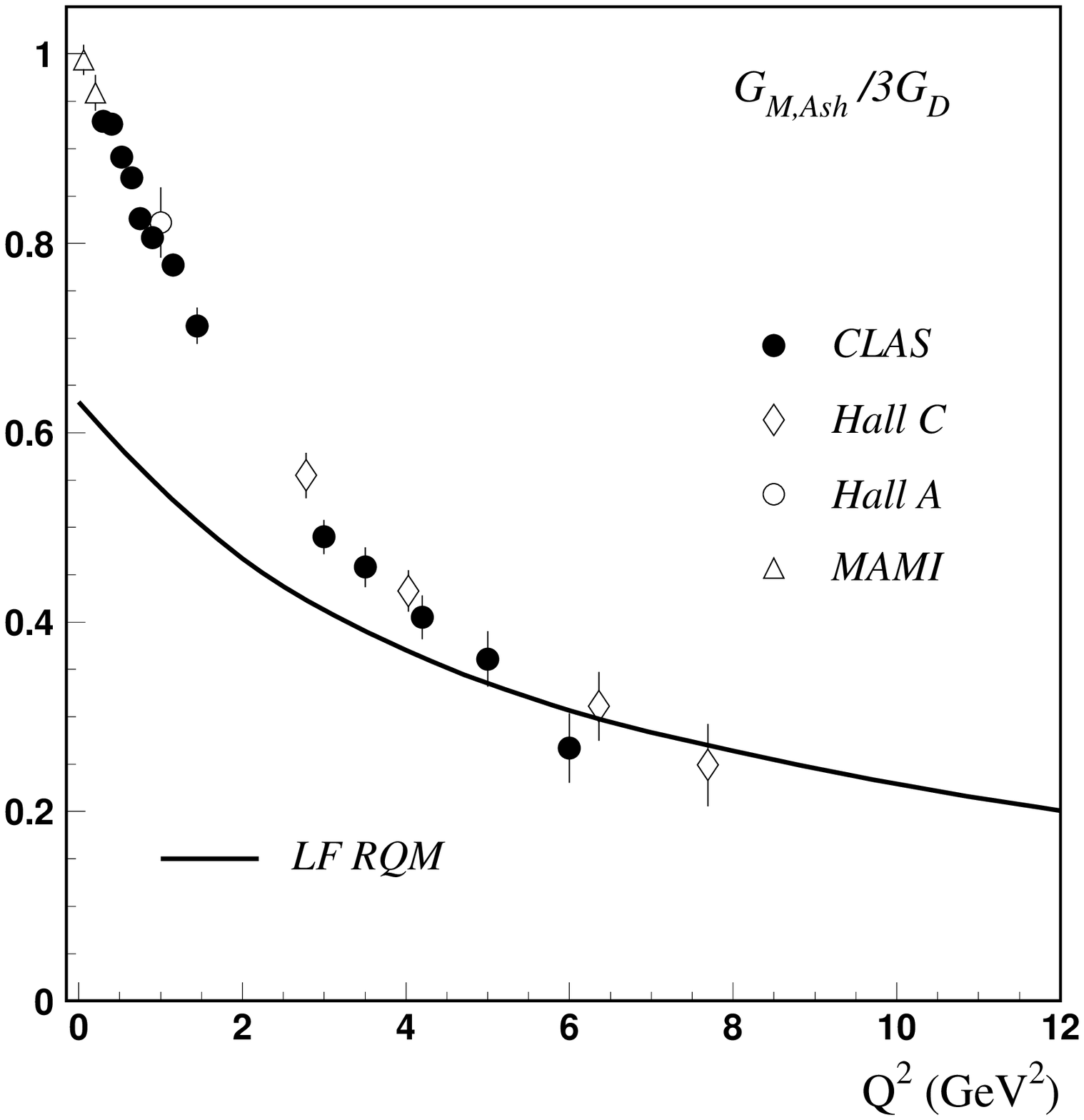}
\includegraphics[width=6.1cm]{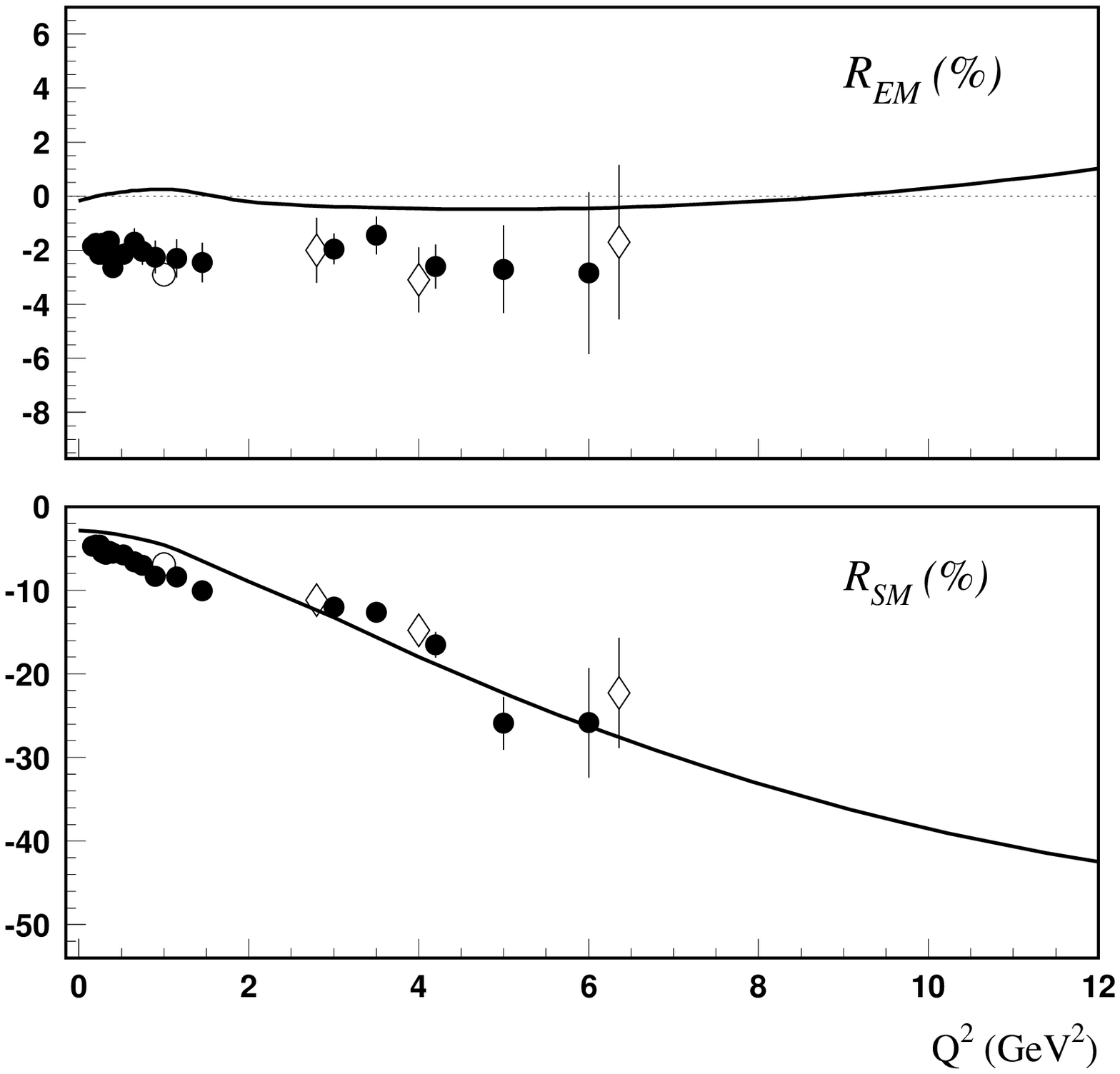}
\caption{\small
The form factor $G_{M,Ash}(Q^2)$ 
for the  $\gamma^* p \rightarrow~\Delta(1232)\frac{3}{2}^+$
transition relative to $3G_D$: $G_D(Q^2)=1/(1+Q^2/0.71GeV^2)$,
and the ratios $R_{EM}$ and $R_{SM}$.
The solid curves are the LF RQM predictions.
The solid circles are the data extracted from the CLAS pion electroproduction data
\cite{Aznauryan2009}. The results from other experiments are:
rhombuses \cite{Frolov,Vilano}, open circle \cite{KELLY},
and open triangles \cite{Stave2006,Sparveris2007,Stave2008}.}
\label{delta}
\end{figure}

\begin{figure}
\includegraphics[width=11.8cm]{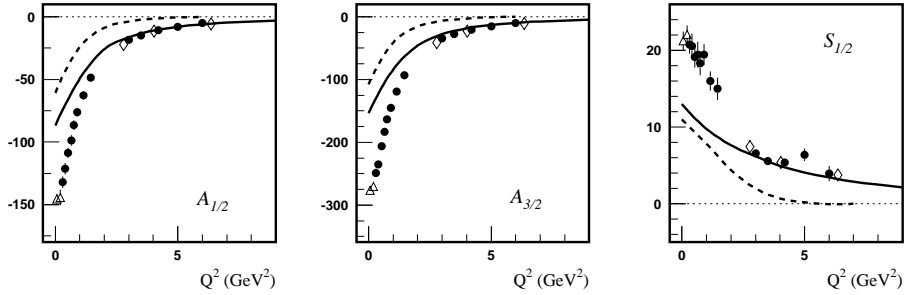}
\caption{\small
The $\gamma^*p\rightarrow \Delta(1232)\frac{3}{2}^+$
transition helicity amplitudes (in units of $10^{-3}~{\rm GeV}^{-1/2}$).
The solid curves correspond to the LF RQM predictions,
the dashed curves present inferred meson-baryon contributions.
The legend for the data is as for Fig. \ref{delta}.}
\label{delta_ampl}
\end{figure}

\begin{figure}
\begin{center}
\includegraphics[width=4.9cm]{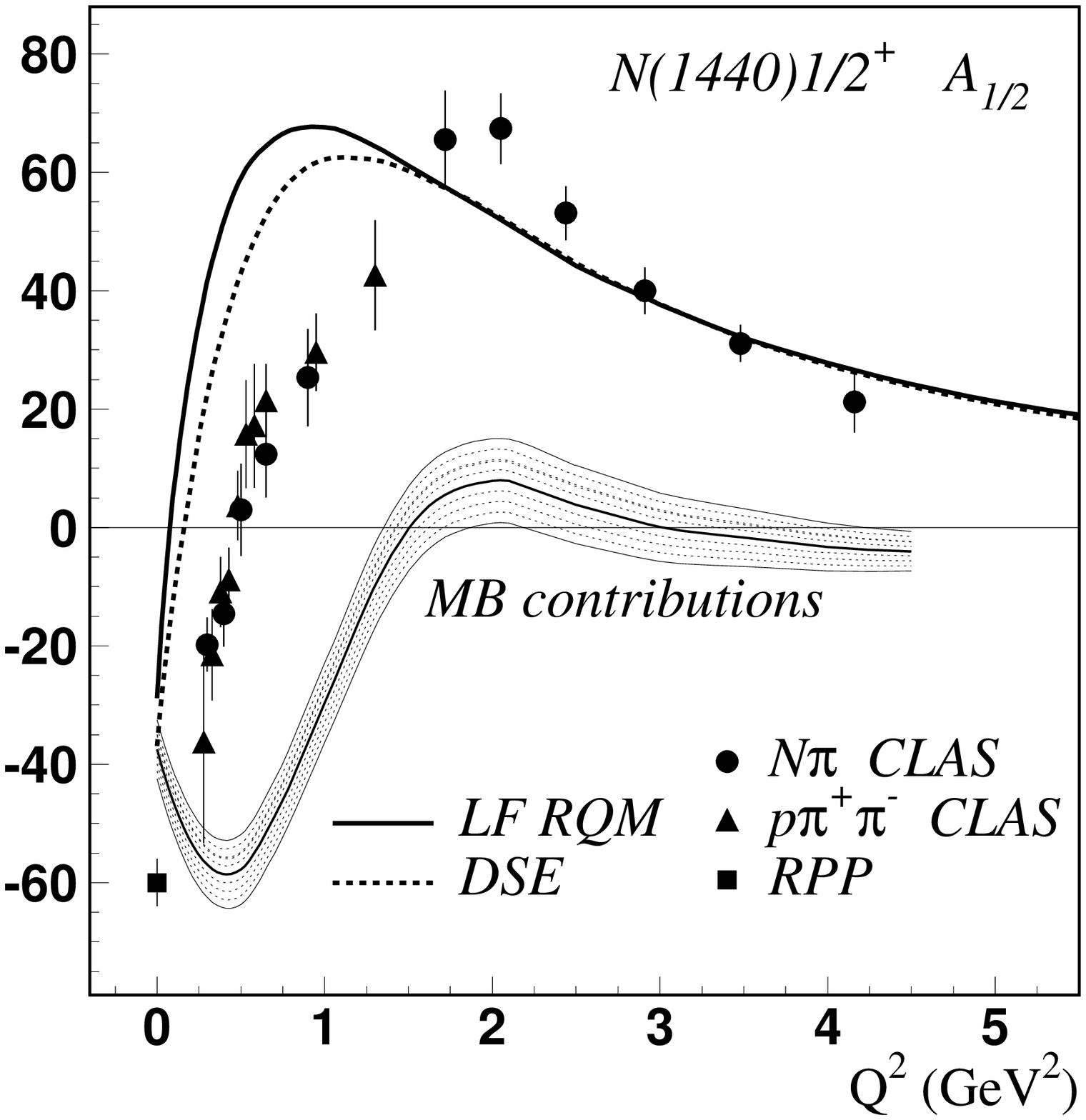}
\includegraphics[width=4.9cm]{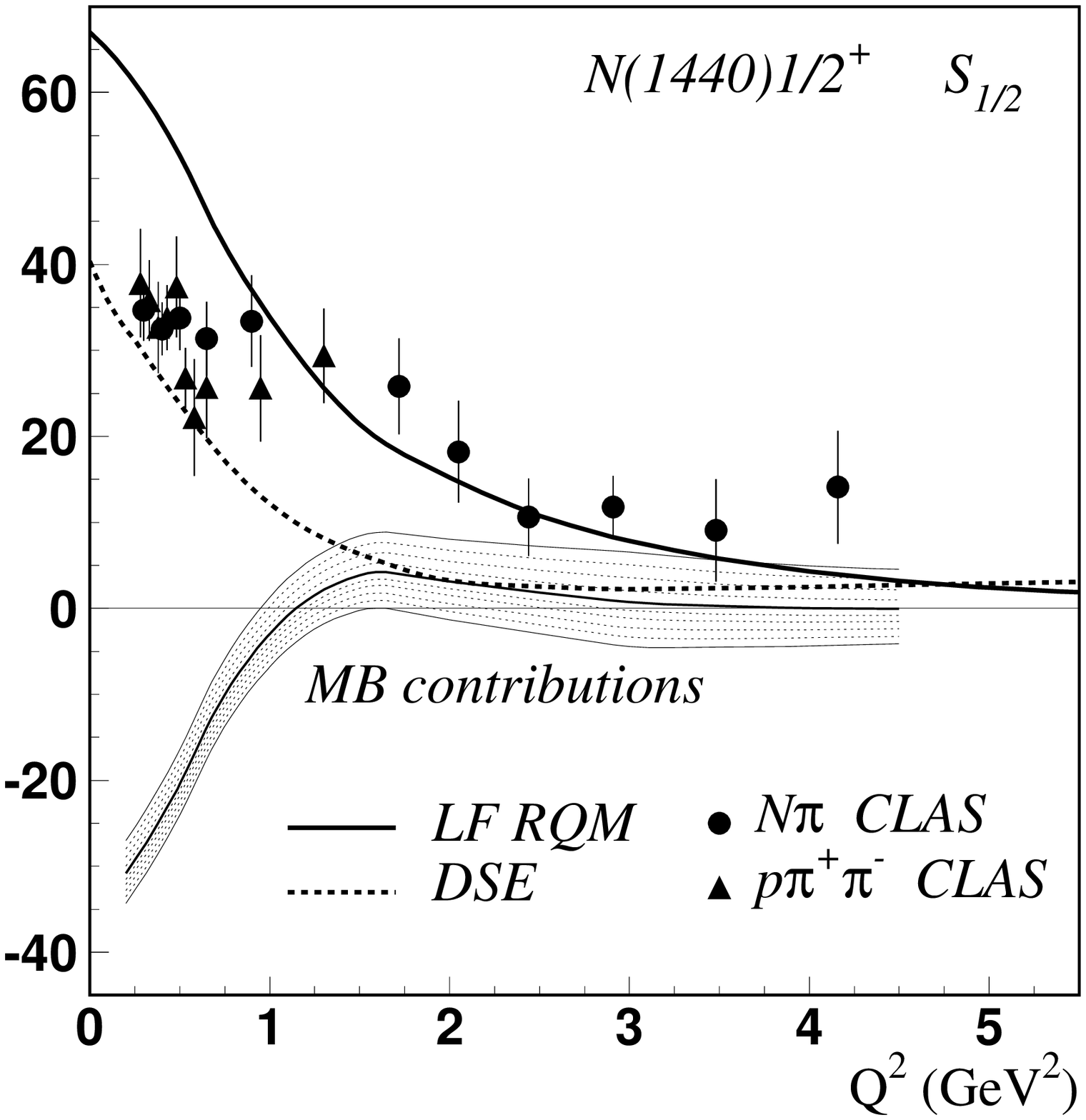}
\end{center}
\caption{\small
The $\gamma^*p\rightarrow N(1440)\frac{1}{2}^+$
transition helicity amplitudes (in units of $10^{-3}~{\rm GeV}^{-1/2}$).
The solid curves are the LF RQM predictions,
the dotted curves present
the results obtained within Dyson-Schwinger equations \cite{Roberts,Roberts1}.
The solid circles and triangles are the amplitudes extracted
from the CLAS data on $\pi N$ \cite{Aznauryan2009} and $\pi^+\pi^- p$
\cite{Mokeev,Mokeev1} electroroduction off the proton.
The solid box at $Q^2=0$ is the RPP estimate \cite{RPP}.}
\label{roper}
\end{figure}

\begin{figure}
\includegraphics[width=3.9cm]{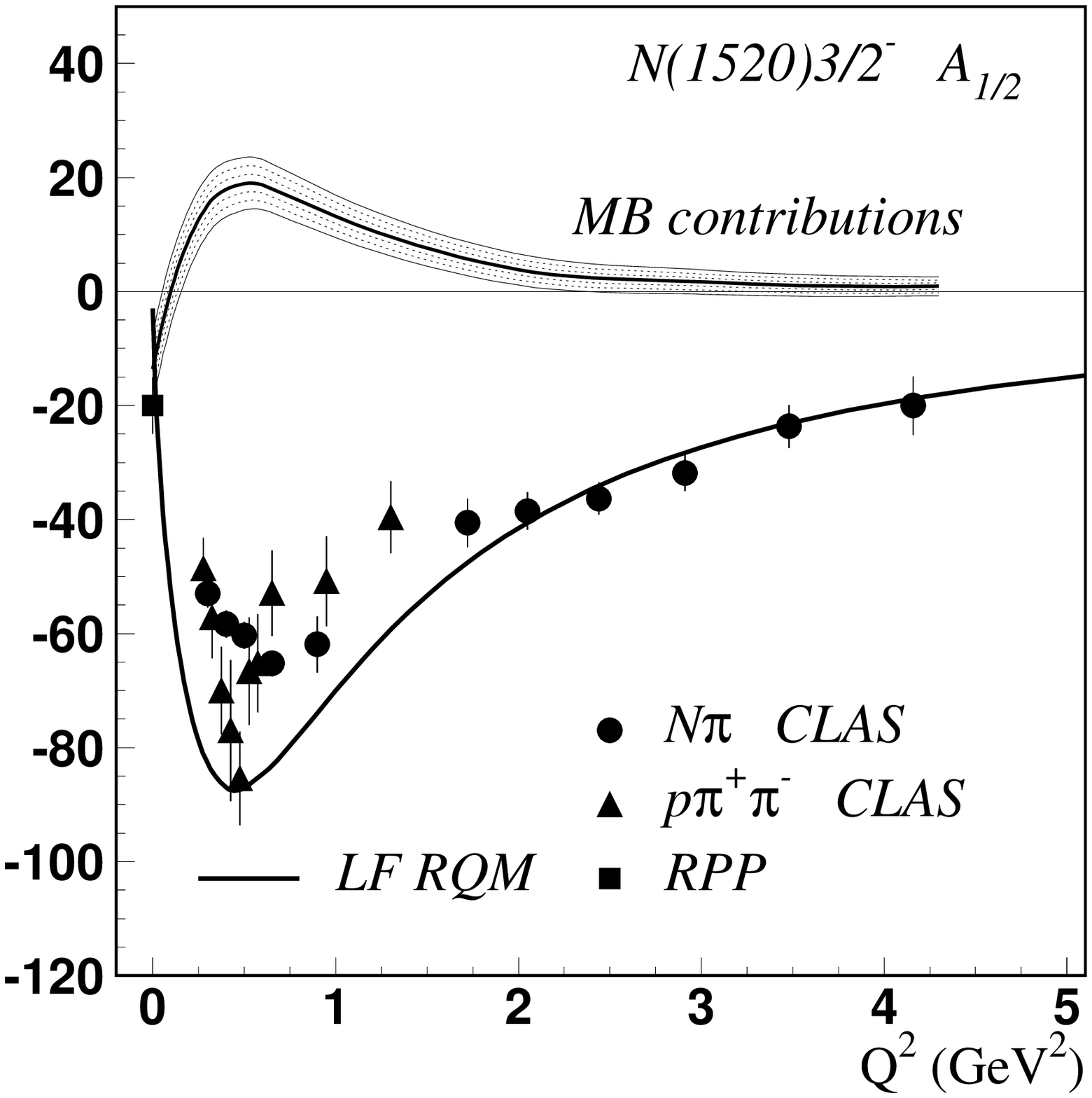}
\includegraphics[width=3.9cm]{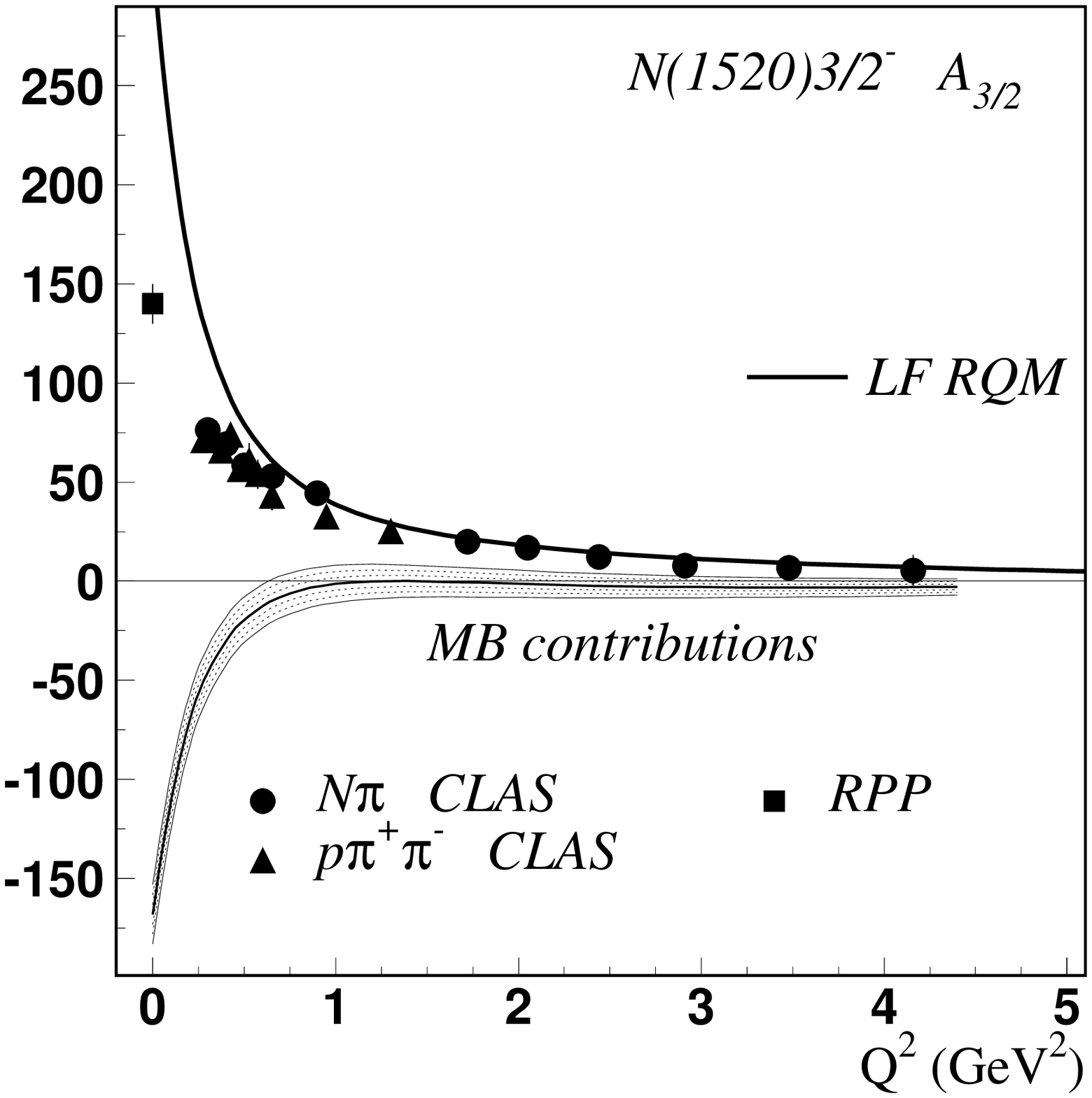}
\includegraphics[width=3.9cm]{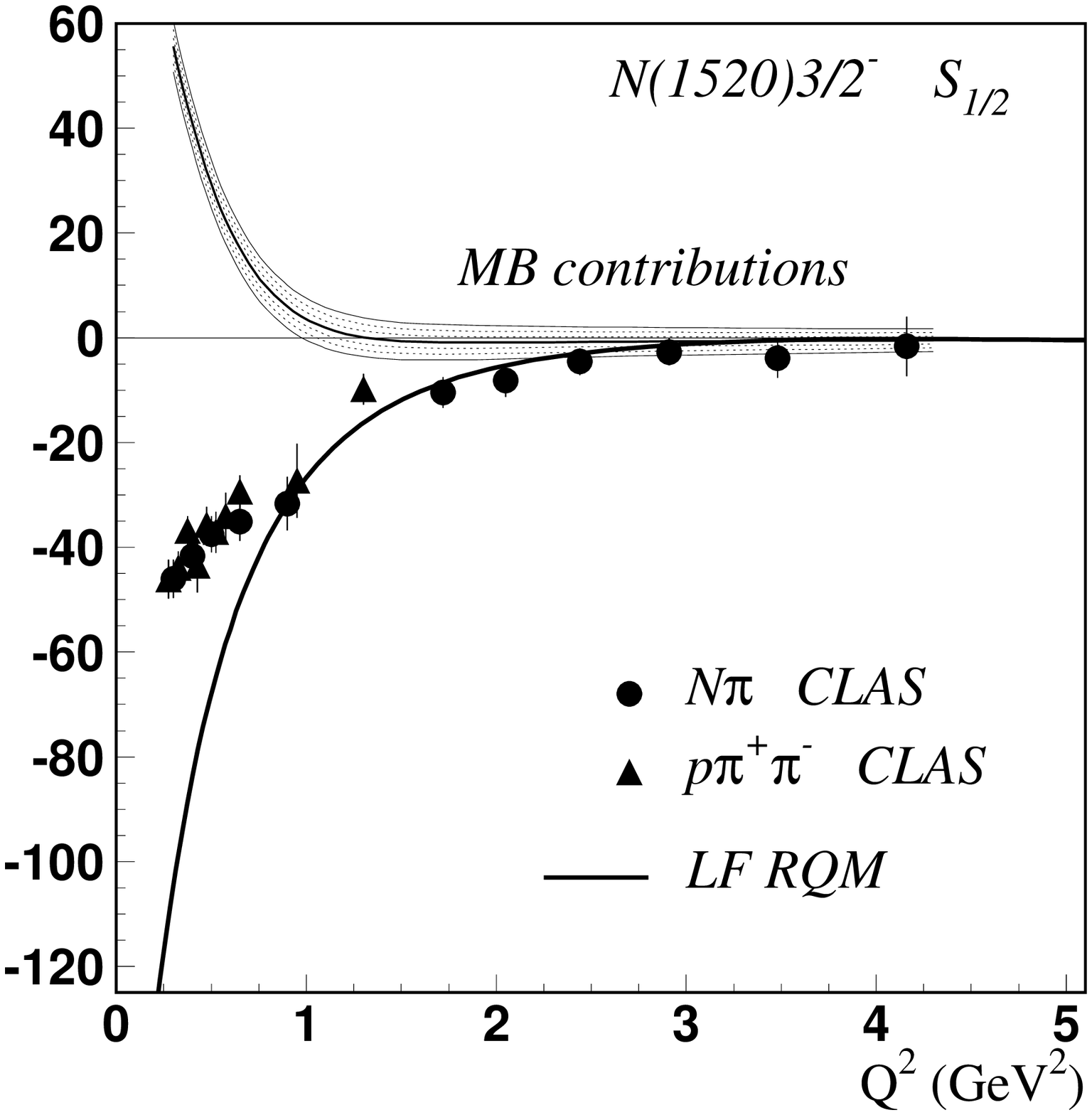}
\caption{\small
The $\gamma^*p\rightarrow N(1520)\frac{3}{2}^-$
transition helicity amplitudes (in units of $10^{-3}~{\rm GeV}^{-1/2}$).
The legend for the curves and data is as for Fig. \ref{roper}.}
\label{d13}
\end{figure}

\begin{figure}
\begin{center}
\includegraphics[width=4.9cm]{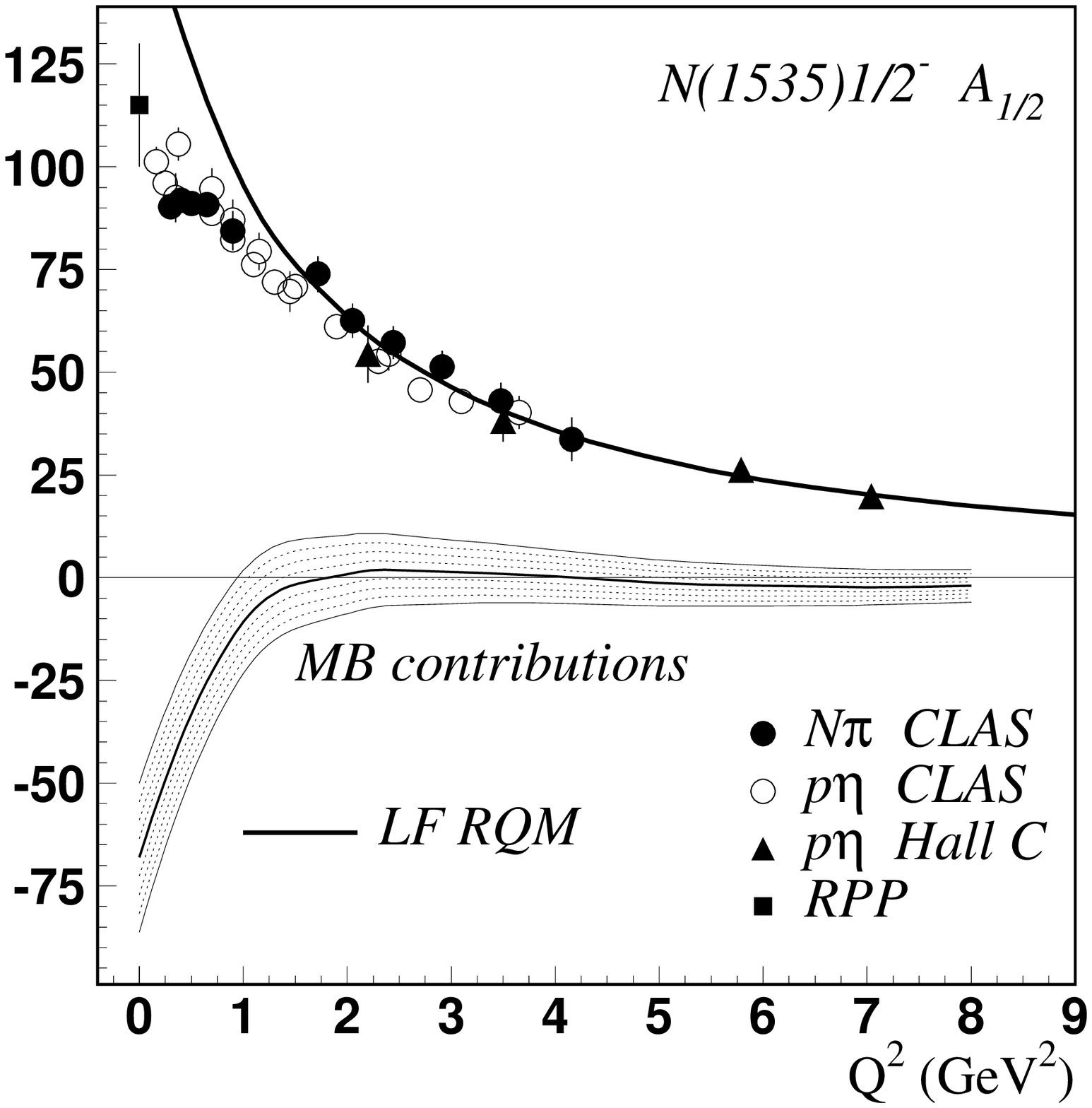}
\includegraphics[width=4.9cm]{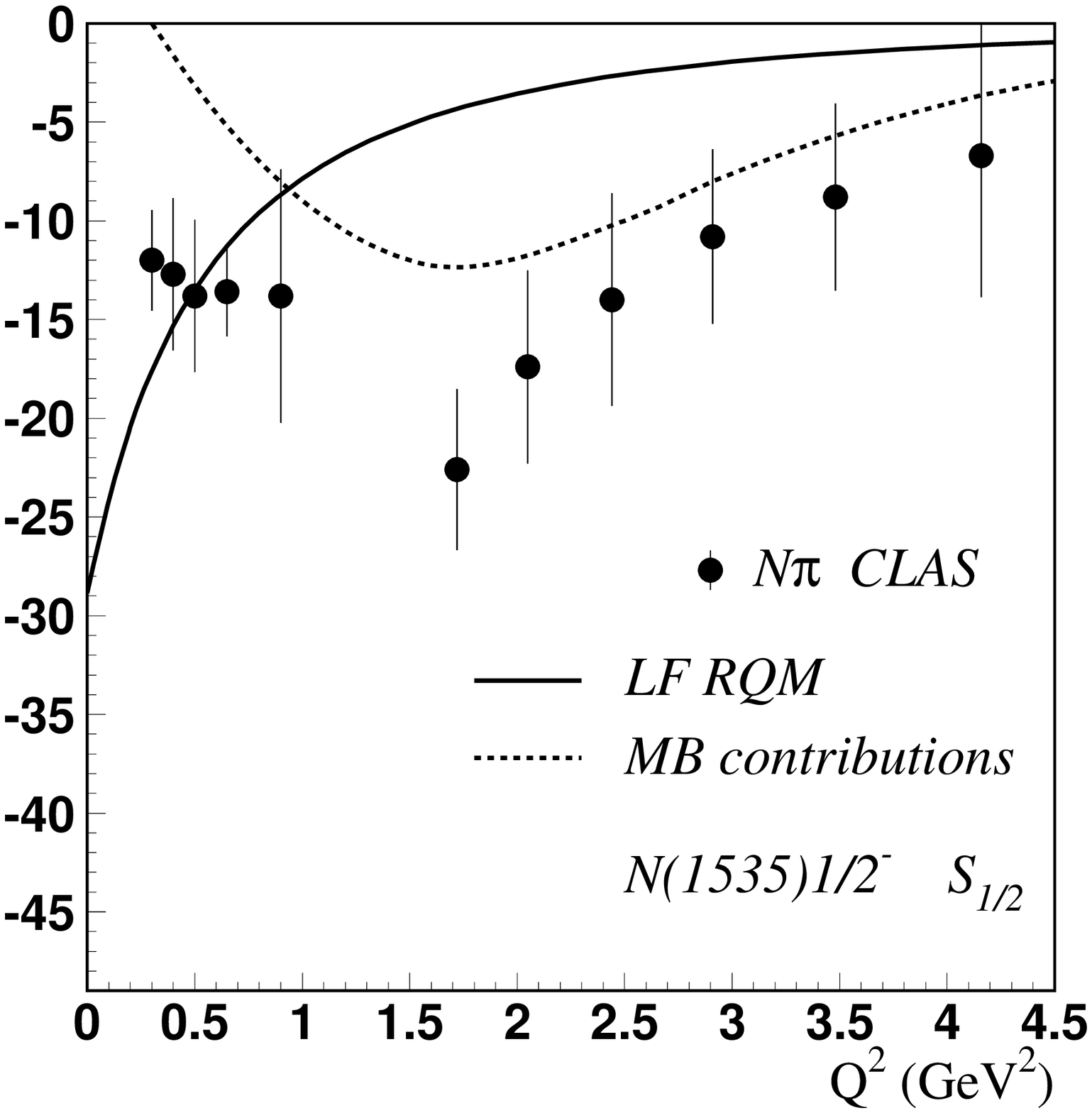}
\end{center}
\caption{\small
The $\gamma^*p\rightarrow N(1535)\frac{1}{2}^-$
transition helicity amplitudes (in units of $10^{-3}~{\rm GeV}^{-1/2}$).
The solid curves are the LF RQM predictions.
The solid circles are the amplitudes extracted from CLAS
pion electroproduction data \cite{Aznauryan2009},
the open circles are the amplitudes extracted from CLAS $\eta$ electroproduction
data \cite{Thompson,Denizli}, the solid triangles are the amplitudes
extracted from JLab/Hall C $\eta$ electroproduction data \cite{Armstrong,Dalton}.
The solid box at $Q^2=0$ is the RPP estimate \cite{RPP}.}
\label{s11}
\end{figure}

\begin{figure}
\begin{center}
\includegraphics[width=3.9cm]{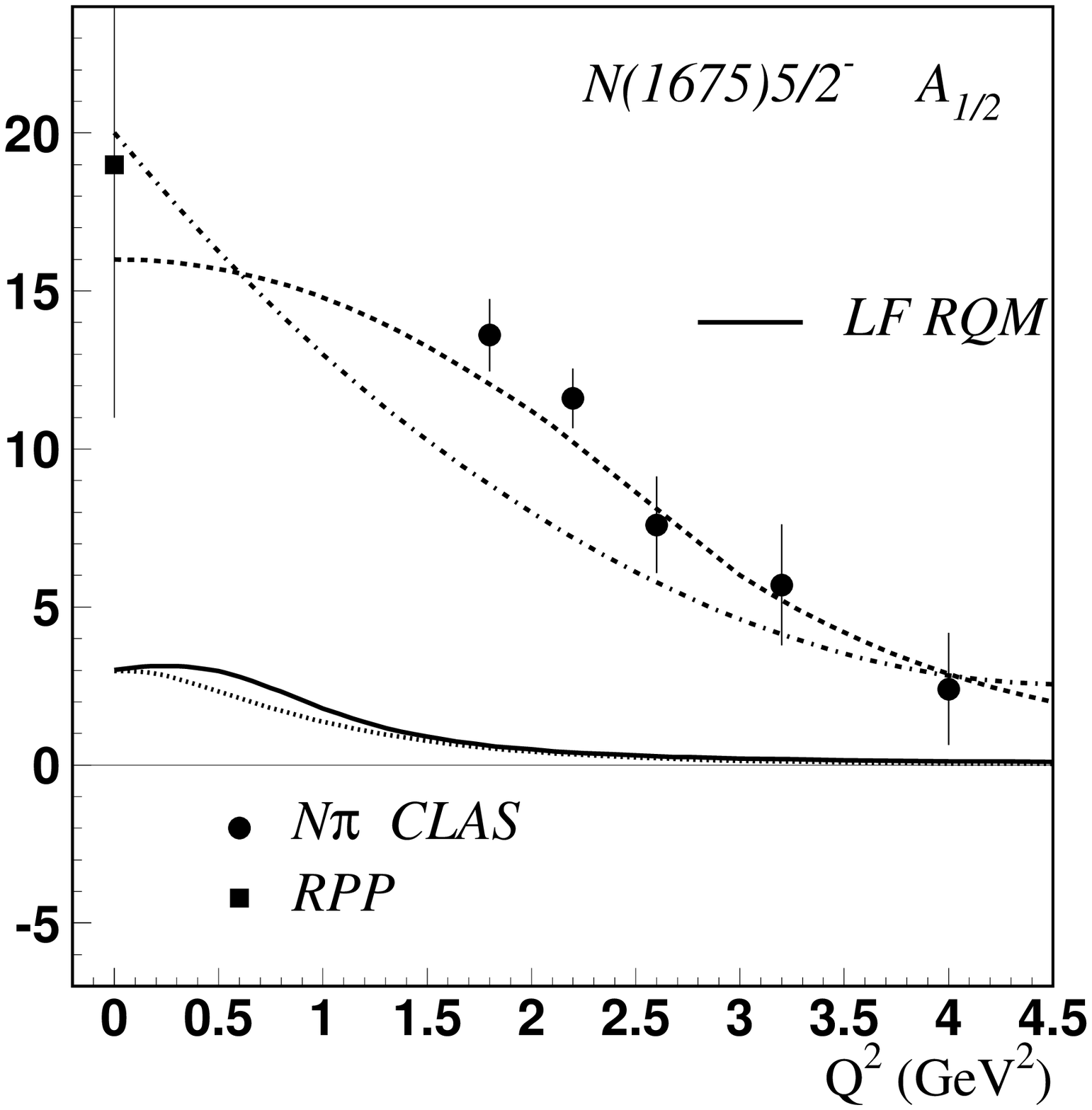}
\includegraphics[width=3.9cm]{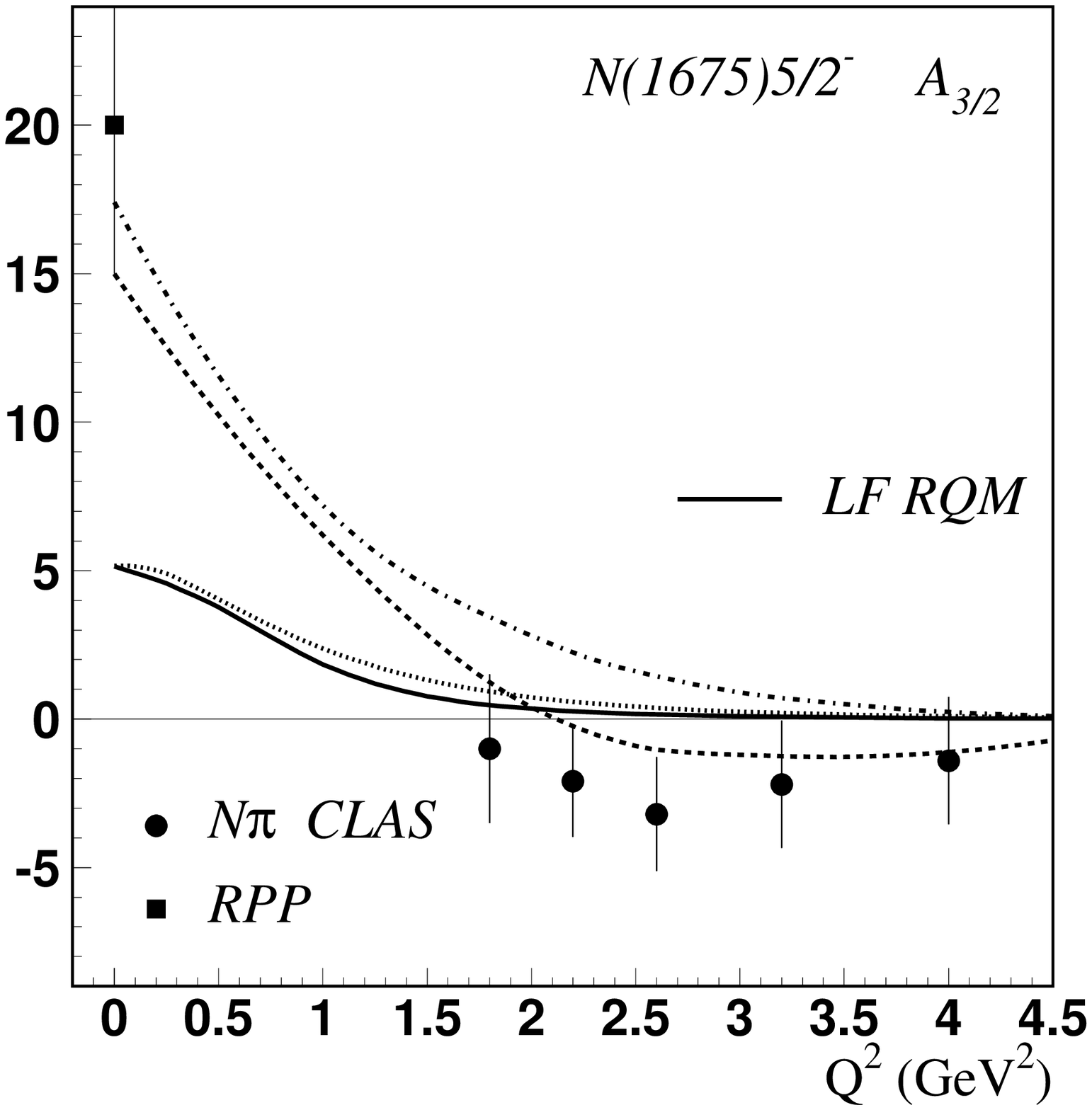}
\includegraphics[width=3.9cm]{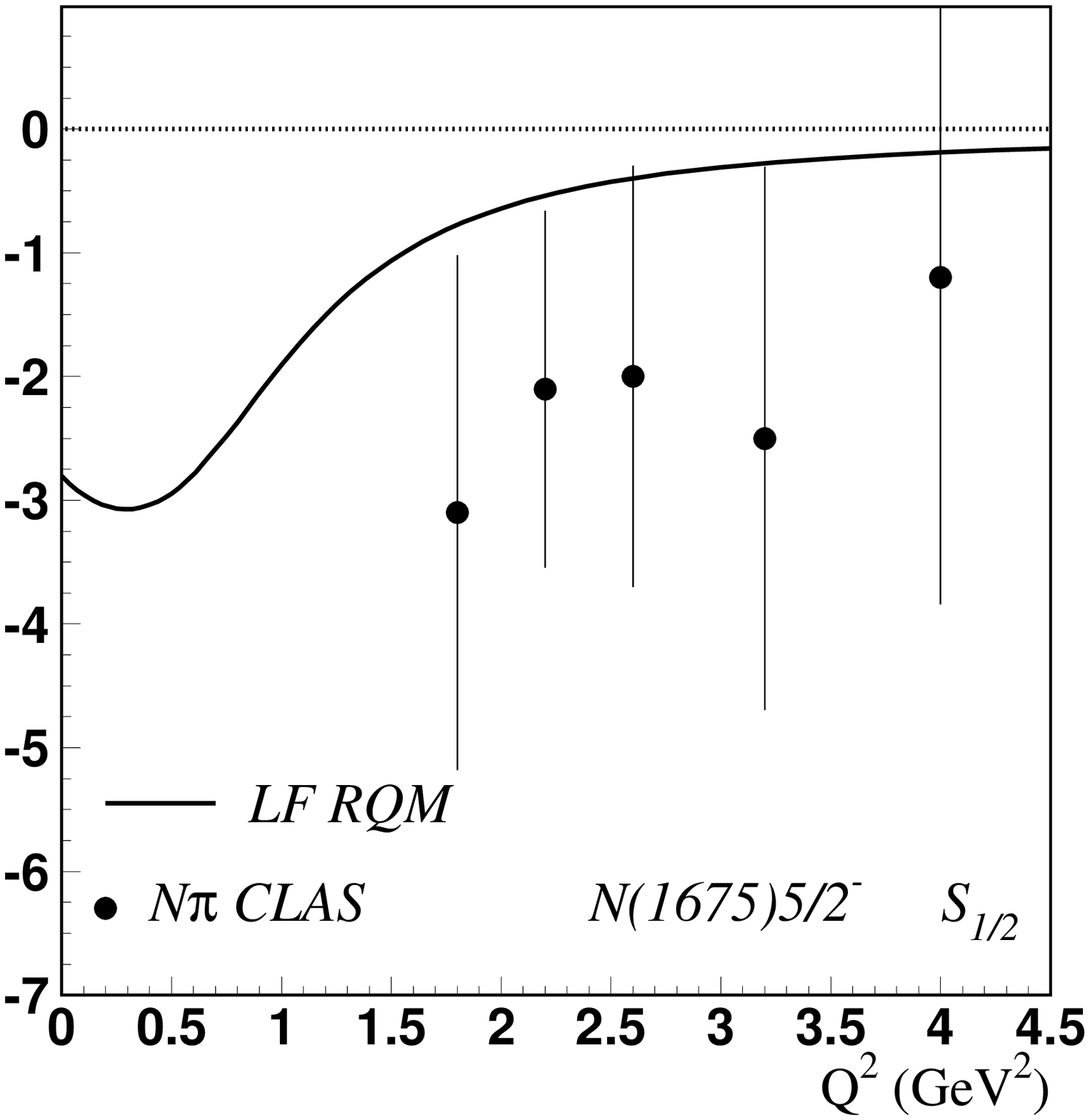}
\end{center}
\caption{\small
The $\gamma^*p\rightarrow N(1675)\frac{5}{2}^-$
transition helicity amplitudes (in units of $10^{-3}~{\rm GeV}^{-1/2}$).
The solid curves are the LF RQM predictions,
the dashed curves present inferred meson-baryon contributions.
the dashed-dotted curves are
absolute values of the predicted
meson-baryon contributions from the dynamical coupled-channel
approach of Ref. \cite{Lee1}, and
the dotted curves correspond to quark model
predictions of Ref. \cite{San}.
The solid circles are the amplitudes extracted from CLAS
pion electroproduction data \cite{Aznauryan2009}.
The solid box at $Q^2=0$ is the RPP estimate \cite{RPP}.}
\label{d15}
\end{figure}

We present the results for the electroexcitation
of the resonances in terms of transition helicity amplitudes
(Figs. \ref{delta_ampl}-\ref{d13},\ref{s11},\ref{d15}).
The exception is made for the $\Delta(1232)\frac{3}{2}^+$.
For this resonance in Fig. \ref{delta} we present also the predictions for
the $\gamma^* p \rightarrow~\Delta(1232)\frac{3}{2}^+$
magnetic-dipole transition form factor and the ratios
$R_{EM}\equiv ImE^{3/2}_{1+}/ImM^{3/2}_{1+}$,
$R_{SM}\equiv ImS^{3/2}_{1+}/ImM^{3/2}_{1+}$, as
these observables are commonly used to present the results on the
$\Delta(1232)\frac{3}{2}^+$ extracted from experimental data on
the electroproduction of pions on nucleons.
From Fig. \ref{delta} it is seen that the 
predictions for the $3q$ contribution to the
$\gamma^* N \rightarrow \Delta(1232)\frac{3}{2}^+$
magnetic-dipole form factor are within limits
obtained in the dynamical reaction model
\cite{Lee1,Lee}, where the bare contribution, that can be associated with
the $3q$ contribution, gives at $Q^2=0$
about $40-70\%$ of the total magnetic-dipole
form factor.
For the ratios $R_{EM}$ and $R_{SM}$ one conclude
that $3q$ core contribution to
$R_{EM}$ will stay small up to $Q^2 = 12$~GeV$^2$, while
for $R_{SM}$ it grows in agreement with experimental data
and will continue to grow up $Q^2 = 12$~GeV$^2$. 

The CLAS measurements made possible, for the first time, the determination
of the electroexcitation amplitudes of the Roper resonance  
$N(1440)\frac{1}{2}^+$ on the proton in wide range of $Q^2$ 
\cite{Aznauryan2008,Aznauryan2009}. Comparison of the data with
the LF RQM predictions \cite{Aznauryan2007,Capstick1995}
provided strong evidence for the identification of 
the $N(1440)\frac{1}{2}^+$ as a predominantly first radial excitation
of the nucleon \cite{Aznauryan2008,Aznauryan2009,review}. 
This conclusion was based on the description of the 
following specific features in the extracted
$\gamma^* p \rightarrow N(1440)\frac{1}{2}^+$ amplitudes:
(1) the specific behavior of the transverse
amplitude $A_{1/2}$, which being large and negative at $Q^2=0$,
becomes large and positive at $Q^2\simeq 2~$GeV$^2$, and then drops slowly
with $Q^2$; (2) the positive relative sign between the longitudinal $S_{1/2}$ 
and transverse $A_{1/2}$ amplitudes above $Q^2\simeq 1~$GeV$^2$;
(3) the common sign of the amplitudes $A_{1/2},S_{1/2}$ 
extracted from the data on $\gamma^* p\rightarrow \pi N$, that includes the
signs from  the $\gamma^* p \rightarrow N(1440)\frac{1}{2}^+$
and $N(1440)\frac{1}{2}^+\rightarrow \pi N$ vertices. 
All these features
are described by the LF RQM 
\cite{Aznauryan2007,Aznauryan2012,Capstick1995} assuming that 
the $N(1440)\frac{1}{2}^+$ is the first radial excitation
of a three-quark ($3q$) ground state. 

\begin{figure}
\begin{center}
\includegraphics[width=6.0cm]{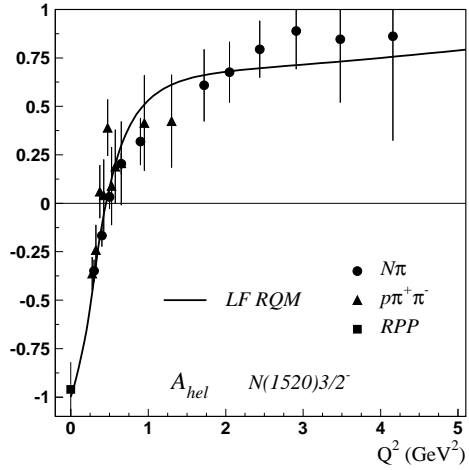}
\end{center}
\caption{\small
The helicity asymmetry $A_{hel}\equiv (A^2_{1/2}-A^2_{3/2})/(A^2_{1/2}+A^2_{3/2})$ 
for the $\gamma^*p\rightarrow N(1520)\frac{3}{2}^-$
transition.
The legend for the curves and data is as for Fig. \ref{roper}.}
\label{d13_asym}
\end{figure}

In Fig. \ref{d13_asym}, we present the comparison of the LF RQM
predictions with experimental data for the helicity
asymmetry $A_{hel}\equiv (A^2_{1/2}-A^2_{3/2})/(A^2_{1/2}+A^2_{3/2})$
in the $\gamma^*p\rightarrow N(1520)\frac{3}{2}^-$
transition. The data, as well LF RQM predictions, show the rapid
helicity switch from the dominance of the $A_{3/2}$ amplitude
at the photon point to the dominance of $A_{1/2}$ at $Q^2~ >~ 1~$GeV$^2$.
Such behavior was predicted by the nonrelativistic 
quark models with harmonic oscillator potential \cite{Close,Isgur}. 
It is reproduced also by the LF RQM.

The approximation
of the single quark transition model 
\cite{Hey_Weyers,Babcock_Rosner,Cottingham,SQTM}
leads to selection rules, which for
the resonance $N(1675)\frac{5}{2}^-$ result
in the suppression of the amplitudes
$A_{1/2}(Q^2)$ and $A_{3/2}(Q^2)$ on the proton.
According to our results,
relativistic effects violate this suppression weakly,
and we expect that experimental values of these amplitudes
should be dominated by the meson-baryon contributions.
In contrast with the proton, the quark core contributions to the
electroexcitation amplitudes on the neutron
for the $N(1675)\frac{5}{2}^-$ are not suppressed and
are predicted to be large.
In both cases, for the proton and neutron,
similar predictions have been
obtained in the quark model of Ref.~\cite{San}.

The meson-baryon contributions presented in
Figs. \ref{delta_ampl}-\ref{d13},\ref{s11},\ref{d15} are inferred from the difference
of the LF RQM predictions and the data.
Most of these contributions
have a clear peak at $Q^2=0$, except for the $A_{1/2}(Q^2)$ amplitude of
$N(1520)\frac{3}{2}^-$ and for the $S_{1/2}(Q^2)$ amplitude
of $N(1535)\frac{1}{2}^-$. Such pronounced peaks are also characteristic
for the meson cloud contributions in the
coupled-channels analyses
\cite{Lee1,Lee2}.
Concerning  the $A_{1/2}(Q^2)$ amplitude for
the $N(1520)\frac{3}{2}^-$, we mention that in all
coupled-channels analyses
the results for the meson
cloud contribution are by order of magnitude
and $Q^2$ dependence very similar to our result.

At the photon point $Q^2=0$, the inferred meson-baryon
contributions 
for the $N(1400)\frac{1}{2}^+$, 
$N(1520)\frac{3}{2}^-$, $N(1535)\frac{1}{2}^-$, and
$N(1675)\frac{5}{2}^-$ 
can be found on the proton and neutron
using the RPP (Review of Particle Physics) estimates \cite{RPP}.
According to our results presented in Table \ref{cloud}, for these resonances
the inferred meson-baryon contributions at the photon point 
are dominated by the isovector component.

\begin{acknowledgements}
This work was supported by
the U.S. Department of Energy, Office of Science,
Office of Nuclear Physics, under Contract
No. DE-AC05-06OR23177, and the National Science Foundation, State
Committee of Science of the Republic of Armenia, Grant No. 15T-1C223.
\end{acknowledgements}

\begin{table}
\caption{
Transverse transition helicity
amplitudes at $Q^2=0$ for several states
for proton and neutron (in units of $10^{-3}~{\rm GeV}^{-1/2}$).
The first two columns show the RPP estimates \cite{RPP}.
Columns 3 and 4 show the inferred meson-baryon contributions obtained by
subtracting the values obtained in the LF RQM and those from experimental data.
The quoted uncertainties are from the experimental estimates.}
\label{cloud}       
\begin{center}
\begin{tabular}{|c|c|}
\hline
&\\   
Resonance&
$A_{1/2}~~~~~~~~~~A_{3/2}
~~~~~~~~~A_{1/2}~~~~~~~~~~A_{3/2}$\\
&\\
&$~~~~~$exp. \cite{RPP}$
~~~~~~~~~~~~~~~~~$exp$~-~$LF RQM\\
&\\
\hline
&\\   &proton\\
&\\
$N(1440){\frac{1}{2}}^+$&$-60\pm 4$~~~~~~~~~~~~~~~~~~~~~~~~$-31\pm 4~~~~~~~~~~~~~~~~~$\\
&\\
$N(1520){\frac{3}{2}}^-$&$-20\pm 5$~~~~~~~~$140\pm 10$ ~~~~$-17\pm 5$ ~~~ $-174\pm 10$\\
&\\
$N(1535){\frac{1}{2}}^-$&$115\pm 15$~~~~~~~~~~~~~~~~~~~~~~~$-54\pm 15~~~~~~~~~~~~~~~~$\\
&\\
$N(1675){\frac{5}{2}}^-$&$~~19\pm 8$~~~~~~~~~~$20\pm 5$~~~~~~~~~$16\pm 8$ ~~~~~~~ $15\pm 5$\\
&\\   
&\\
\hline
&\\   &neutron\\
&\\
$N(1440){\frac{1}{2}}^+$&$~~~40\pm 10~~~~~~~~~~~~~~~~~~~~$
$~12\pm 10~~~~~~~~~~~~~~~~$\\
&\\
$N(1520){\frac{3}{2}}^-$&$-50\pm 10$~~~$-115\pm 10$
~~~~$20\pm 10$~~~~$131\pm 10$\\
&\\
$N(1535){\frac{1}{2}}^-$&$~-75\pm 20~~~~~~~~~~~~~~~~~~~~$
$87\pm 20~~~~~~~~~~~~~~~~~$\\
&\\
$N(1675){\frac{5}{2}}^-$&$-60\pm 5$~~~~~$-85\pm 10$
~~~~$-13\pm 5$~~~~$-23\pm 10$\\
&\\
\hline
\end{tabular}
\end{center}
\end{table}

\end{document}